\documentclass[aps,prd,twocolumn,nofootinbib,superscriptaddress]{revtex4}
\usepackage{color}
\usepackage{graphicx}
\usepackage{epsfig}

\def\gsim{\:\raisebox{-0.5ex}{$\stackrel{\textstyle>}{\sim}$}\:}

\def\bg#1{\mathchoice
{{\mbox{\boldmath $#1$}}}
{{\mbox{\boldmath $#1$}}}
{{\mbox{\boldmath $\scriptstyle #1$}}}
{\mbox{\boldmath $\scriptscriptstyle #1$}}} 
\def\beq{\begin{equation}}
\def\eeq{\end{equation}}
\def\beqa{\begin{eqnarray}}
\def\eeqa{\end{eqnarray}}
\def\anu{\bar{\nu}}
\def\nue{\nu_e}
\def\anue{\bar{\nu}_e}
\def\numu{\nu_\mu}
\def\anumu{\bar{\nu}_\mu}
\def\nutau{\nu_\tau}
\def\anutau{\bar{\nu}_\tau}
\def\nui{\nu_i}
\def\nux{\nu_x}
\def\anux{\bar{\nu}_x}
\def\Enu{E_\nu}
\def\Enui{E_{\nui}}
\def\Enuiav{\langle\Enui\rangle}
\def\Enuisqav{\langle E^2_{\nui}\rangle}
\def\alphanui{\alpha_{\nui}}
\def\boldq{\mbox{\boldmath$q$}}
\def\tauplus{\tau_+}
\def\tauminus{\tau_-}
\def\gammazero{\gamma_0}
\def\sss{\scriptscriptstyle}

\def\half{\frac{1}{2}}

\def\cm{\,{\rm cm}}

\def\kpc{\,{\rm kpc}}

\def\msun{M_\odot}

\newcommand{\GT}{\rm GT}
\newcommand{\F}{\rm F}
\newcommand{\NO}{\rm NO}
\newcommand{\IO}{\rm IO}
\newcommand{\Fe}{^{56}{\rm Fe}}
\newcommand{\Co}{^{56}{\rm Co}}
\newcommand{\Mn}{^{56}{\rm Mn}}
\newcommand{\Pb}{^{208}{\rm Pb}}

\newcommand{\Bi}{^{208}{\rm Bi}}
\newcommand{\X}{{\rm X}}
\newcommand{\Y}{{\rm Y}}
\newcommand{\etal}{{\it et al. }}

\begin{document}

\title{Detecting supernova neutrinos with iron and lead detectors} 
\author{Abhijit Bandyopadhyay}
\email{ abhijit@rkmvu.ac.in}
\affiliation{\mbox{Ramakrishna Mission Vivekananda University, Belur 
Math, Howrah 711202, India}}
\author{Pijushpani Bhattacharjee}
\email{pijush.bhattacharjee@saha.ac.in}
\affiliation{\mbox{Saha Institute of Nuclear Physics, 1/AF Bidhannagar, 
Kolkata 700064, India}} 
\author{Sovan Chakraborty}
\email{sovan@iitg.ac.in}
\affiliation{\mbox{Department of Physics, Indian Institute of 
Technology - Guwahati, Guwahati 781039, India}}
\affiliation{Max-Planck-Institut f\"{u}r Physik 
(Werner-Heisenberg-Institut),\ F\"{o}hringer Ring 6,\ 80805 
M\"{u}nchen,\ Germany}
\author{Kamales Kar}
\email{kamales.kar@gmail.com}
\affiliation{\mbox{Ramakrishna Mission Vivekananda University, Belur 
Math, Howrah 711202, India}}
\author{Satyajit Saha}
\email{satyajit.saha@saha.ac.in}
\affiliation{\mbox{Saha Institute of Nuclear Physics, 1/AF
Bidhannagar, Kolkata 700064, India}}

\begin{abstract}
Supernova (SN) neutrinos can excite the nuclei of various detector 
materials beyond their neutron emission thresholds through charged 
current (CC) and neutral current (NC) interactions. The emitted 
neutrons, if detected, can be a signal for the supernova event. Here we 
present the results of our study of SN neutrino detection through the 
neutron channel in $\Pb$ and $\Fe$ detectors for realistic neutrino 
fluxes and energies given by the recent Basel/Darmstadt simulations for 
a 18 solar mass progenitor SN at a distance of 10 kpc. We find that, in 
general, the number of neutrons emitted per kTon of detector material 
for the neutrino luminosities and average energies of the different 
neutrino species as given by the Basel/Darmstadt simulations are 
significantly lower than those estimated in previous studies based on 
the results of earlier SN simulations. At the same time, we 
highlight the fact that, although the total number of neutrons produced 
per kTon in a $\Fe$ detector is more than an order of magnitude lower 
than that for $\Pb$, the dominance of the flavor blind NC events in the 
case of $\Fe$, as opposed to dominance of $\nue$ induced CC events in 
the case of $\Pb$, offers a complementarity between the two detector 
materials so that simultaneous detection of SN neutrinos in a $\Pb$ and 
a sufficiently large $\Fe$ detector suitably instrumented for neutron 
detection may allow estimating the fraction of the total $\mu$ and 
$\tau$ flavored neutrinos in the SN neutrino flux and thereby 
probing the emission mechanism as well as flavor oscillation scenarios 
of the SN neutrinos.  

\end{abstract}
\maketitle

\section{Introduction}\label{sec:intro}
Detection of the neutrinos emitted during core collapse supernova (SN) 
explosion events is important for two reasons. Firstly, these neutrinos 
carry information about the core of the exploding star whereas no other 
particle or radiation can come out of that very high density region. 
Secondly, the properties of neutrinos like their mass hierarchy and 
flavor mixing, and their charged and neutral current interactions with 
matter inside the supernova may leave some imprints on the number of 
neutrinos detected and their temporal structure 
\cite{Raffelt:1999tx,duan-09}, thereby allowing those neutrino 
properties as well as the core collapse supernova explosion mechanism to 
be probed. Because of these reasons a 
number of detectors capable of detecting SN neutrinos have 
come into operation during the past twenty five years or so after the 
detection of neutrinos from supernova 1987A located in 
the Large Magellanic Cloud at a distance of $\sim 
50\kpc$~\cite{Kamioka-II,IMB,Baksan,Mont-Blanc}. For a recent review 
of the capabilities and detection methods of currently operating as well 
as near-future and proposed future SN neutrino detectors, see 
Ref.~\cite{Scholberg:2012id}.

In this paper we study the possibility of detection of SN neutrinos with 
iron or lead as detector materials through detection of neutrons 
emitted from the nuclei excited by the SN neutrinos. The use of 
such heavy-nuclei materials for detection of SN neutrinos through the 
neutron channel has been discussed by a number of authors in the 
past~\cite{SN-nu-heavy-nuclei-detectors,Kolbe-Langanke-01,Engel-etal-03}.   
In general, neutron rich nuclei offer good sensitivity to $\nue$'s 
through charged current (CC) process $\nue+n\to p+e^-$ in contrast to 
water Cerenkov or organic scintillator based detectors which are 
primarily sensitive to $\anue$'s through the CC inverse beta decay 
process $\anue + p \to n + e^+$. Further, CC cross section for $\nue$ 
interaction with high $Z$ nuclei receives significant enhancement due to 
Coulomb effect on the emitted electron, and correlated nucleon 
effect also amplifies the $\nu$-nucleus cross section relative to 
$\nu$-nucleon cross section as a function of $A$. In particular, 
$\Pb$ --- it being both a highly neutron rich ($N=126$) as well as high 
$Z (=82)$ nucleus --- is considered a good material for detection of the 
$\nue$'s from SN through the CC reaction $\nue + \Pb \to e^- + {\Bi}^*$, 
with the excited ${\Bi}^*$ nucleus ($N=125, Z=83$) subsequently decaying 
by emitting one or more neutrons.    
For a recent detailed study of the effectiveness of $\Pb$ as a SN 
neutrino detector material, done within the context of the
currently operating HALO~\cite{HALO_detector} detector, see 
Ref.~\cite{Vaananen-Volpe-11}. 

A $\Pb$ detector would, of course, also 
be sensitive to all the six $\nu$ and $\bar{\nu}$ species  
including   
$\numu\,,\anumu\,,\nutau\,,$ and $\anutau$ components through
neutral current (NC) interaction $\nu (\anu) + \Pb \to \nu (\anu) + 
\Pb^*$, with the excited $\Pb^*$ nucleus subsequently decaying by 
emitting one or more neutrons. However, the $\nu\,$-$\Pb$ NC 
cross section in the SN neutrino energy range of interest is typically 
a factor of 20 or so smaller than the $\nue\,$-$\Pb$ CC cross 
section~\cite{Kolbe-Langanke-01}, and even considering equal 
contributions from all six $\nu$ plus $\anu$ species, the total number 
of interactions would be expected to be dominated by those due to the 
$\nue$ CC interactions; see, e.g., Ref.~\cite{Vaananen-Volpe-11}. 
Indeed, our calculations below show that the neutrons from 
NC interactions would comprise $\sim$ 20\% or less of all events in a 
$\Pb$ detector. On the 
other hand, a material with $N\approx Z$, such as $\Fe$ ($N=30, Z=26$), 
while being significantly less neutron rich compared to $\Pb$ and thus 
having a $\nue$ CC cross section more than an order of magnitude less 
than that for $\Pb$ in the relevant SN neutrino energy 
range of interest, the flavor blind $\nu\,$-$\Fe$ NC cross section is 
less than 
the corresponding $\nue\,$-$\Fe$ CC cross section only by 
a factor of $\sim$ 4--5. With six species of $\nu$ plus $\anu$ 
contributing roughly equally, the total number of interactions in a 
$\Fe$ detector may be expected to be dominated by NC interactions. 
Indeed, this 
expectation is borne out by our calculations below, which show that 
$\sim$ 60\% or more 
of the total number of neutrons in a $\Fe$ detector would come from NC 
interactions as compared to $\sim$ 20\% or less in a $\Pb$ detector. 
Thus, an appropriately 
large $\Fe$ detector can be a good NC detector for SN neutrinos. In this 
respect, in absence of separate 
identification of the CC events, simultaneous detection of SN 
neutrinos in a $\Pb$ and a $\Fe$ detector, for example, can, in 
principle,  provide an 
estimate of the fraction of the $\numu$, $\anumu$, $\nutau$ and 
$\anutau$ components in the total SN $\nu$ flux, thereby probing the 
emission as well as flavor oscillation scenarios of SN neutrinos.  

Motivated by the above considerations, in this paper we make a 
comparative study of the efficacies of the two materials, $\Fe$  and 
$\Pb$, as detector materials for SN neutrinos. In doing this, differing 
from previous studies, we use the results of the most recent  
state-of-the-art Basel/Darmstadt (B/D) 
simulations~\cite{Basel-Darmstadt-10} for the supernova neutrino fluxes 
and average energies, which typically yield closer fluxes among 
different neutrino flavors and lower average energies compared to those 
given by earlier simulations (see, e.g., 
\cite{Totani-etal-98,Gava-etal-09}). The B/D models are based on 
spherically 
symmetric general relativistic hydrodynamics including spectral
three-flavor Boltzmann neutrino transport. These simulations are
much more realistic compared to the earlier simulations based on  
simple leakage schemes~\cite{Totani-etal-98} without full Boltzmann 
neutrino transport. The lower average 
energies of the different neutrino species in the B/D simulations are 
related to the significantly larger radii of the neutrinospheres of the 
different neutrino species found in the new simulations as compared to 
those in the previous simulations. Indeed, several  
recent investigations with the full Boltzmann transport equation and 
their consecutive upgrades (e.g., Ref.~\cite{Mueller-etal-12ab}) have 
also consistently shown colder neutrino fluxes compared to the earlier 
SN simulations. As a consequence, as we shall see below, our results 
for the number of neutrons
emitted are, in general, significantly lower than those obtained 
previously. 

We note here in passing that recently an additional avenue of 
flavor independent detection of all six species of SN neutrinos has 
opened up with the advent of very low threshold Dark Matter 
(DM) detectors which are primarily designed to detect the 
Weakly Interacting Massive Particle (WIMP) candidates of DM through 
detection of nuclear recoil events caused by the scattering of WIMPs 
from nuclei of the chosen detector materials (see, for example, the 
recent review~\cite{DM-detection-review-16}). Because of their 
capability to detect very low ($\sim$ keV) energy nuclear recoils, such 
detectors would be sensitive to nuclear recoils caused by coherent 
elastic neutrino-nucleus scattering (CE$\nu$NS) of the relatively low 
energy ($\sim$ few MeV) SN neutrinos of all 
flavors~\cite{Horowitz-etal-03} with the cross section for the 
process roughly proportional to $N^2$~\cite{Freedman-74-77}, where 
$N$ is the neutron number of the detector material. 
These DM detectors, being also potential NC detectors of SN neutrinos 
of all flavors, may thus provide important information 
about the SN neutrino flux complementary to those derived from 
conventional (mostly CC) SN neutrino detectors. For recent studies on 
this topic, see, 
e.g., \cite{Sovan-PB-KK-14,Abe-etal-XMASSS-16,Lang-etal-16}. 
This, however, is beyond the scope of the present paper and will not be 
further discussed here. 

The rest of this paper is arranged as follows: In Section 
\ref{sec:ccSN-nu} we briefly describe neutrino emission from 
core collapse supernovae and the basic results of the B/D 
simulations for a typical explosion of a $18\msun$ star. Section  
\ref{sec:xsecs-n-emission-etc} discusses the CC and NC cross sections 
for interaction of neutrinos with lead and iron nuclei and the process 
of emission of neutrons from these nuclei. The number of neutrons 
emitted as a function of the neutrino energy is 
calculated by folding in the one-, two- and three neutron emission 
probabilities (as functions of the excitation energies of the 
nucleus under consideration) with the differential cross section for 
neutrino induced excitation of the nucleus to different excitation 
energies. Section \ref{sec:results} gives the results
for the number of neutrons emitted and makes a comparative study 
between lead and iron as detector materials. 
The paper ends with a summary and conclusions in Section 
\ref{sec:summary}.

\section{Neutrinos from Core Collapse Supernovae} 
\label{sec:ccSN-nu} 
The current understanding of neutrino emission from a typical core 
collapse SN event is that it occurs in three main phases. For the first 
25 -- 30 ms post bounce,  
there is a burst of electron neutrinos ($\nue$) due to rapid  
deleptonization ($p+e^- \to n + \nue$) in the core. This neutronization 
burst phase is followed by the accretion phase, lasting for a few
hundred ms, when neutrinos and antineutrions of all flavors are emitted. 
These accretion phase neutrinos are powered by the gravitational energy 
released by the in-falling material accreting on to the central object. 
Finally, at the end of the accretion phase the shock (responsible for 
the SN event) breaks through the mantel of the star and the cooling 
phase ensues when neutrinos and antineutrinos of all three species of 
neutrinos are emitted. The cooling phase is the longest one, 
lasting for about 10 seconds, during which the bulk of the total 
explosion energy is emitted in the form of neutrinos. The neutrino 
production in these different phases depends on different 
nuclear processes, the energetics of the phase and size of the 
progenitor core. Thus the luminosities and average energies of these 
neutrinos vary from one phase to another. 
\begin{figure*}[!]
\epsfig{file=Fig_lum-avEnergy-timeprofile.eps,width=0.9\columnwidth}\hskip0.8cm 
\epsfig{file=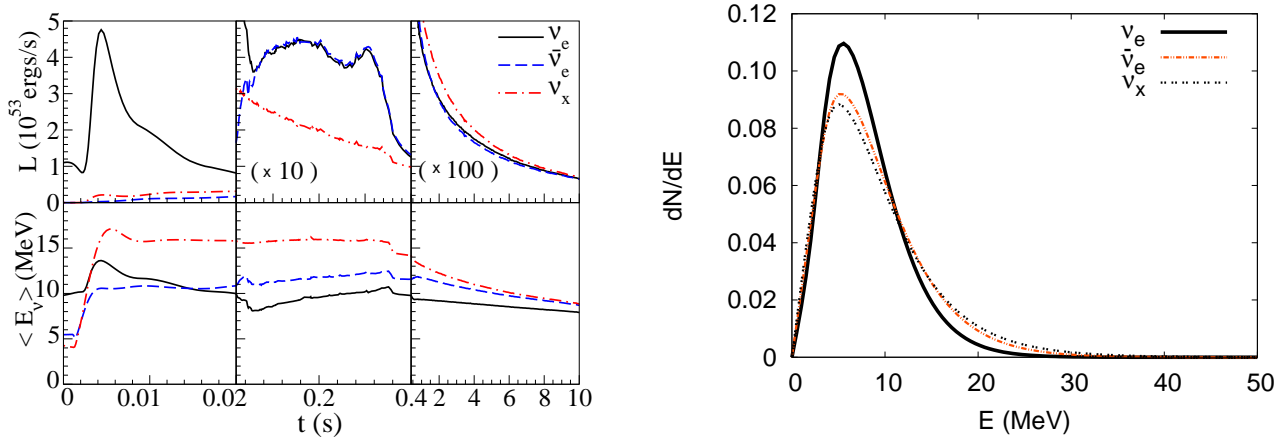,width=\columnwidth}
\caption{Left: Temporal profiles of the neutrino luminosity
(upper three panels) and average energy of the neutrinos (lower three 
panels) corresponding to the neutronization phase, accretion phase and 
cooling phase (from left to right, respectively) for different neutrino 
flavors, obtained from the results of the Basel/Darmstadt 
simulation~\cite{Basel-Darmstadt-10} of a 18 $M_\odot$ progenitor SN 
(From Ref.~\cite{Sovan-PB-KK-14}). 
Right: The normalized time-averaged energy spectra of the neutrinos of 
different flavors.   
} 
\label{fig:lum-avEnergy-timeprofile-spectra}
\end{figure*}

To estimate the total neutrino flux as a function of energy on a 
detector at Earth, one needs the energy spectra of the emitted neutrinos 
and their luminosities in the different phases. The number of primary 
(i.e., at the neutrinosphere) neutrinos of flavor $i$ ($\nu_i\equiv 
\nue\,, \anue\,, \nux\,, \anux$, with $\nux$ representing $\numu$ or 
$\nutau$) emitted per unit time per unit energy is in general written 
as 
\beq
F^0_{\nui} (t,\Enu) = \frac{L_{\nui} (t)}{\Enuiav 
(t)}\varphi_{\nui}(\Enu,t)\,,
\label{eq:Fnuzero}
\eeq
where $L_{\nui} (t)$ and $\Enuiav (t)$ are the time dependent 
luminosity and average energy of the emitted neutrinos of flavor $\nui$, 
and $\varphi_{\nui}(\Enu,t)$ is the instantaneous normalized energy 
spectrum ($\int \varphi_{\nui}(\Enu,t)d\Enu = 1$), which can be 
parametrized as \cite{Keil:2002in}
\beqa
\varphi_{\nui}(\Enu,t) & = \frac{1}{\Enuiav (t)}
\frac{\Big(1+\alphanui (t)\Big)^{1+\alphanui (t)}}{\Gamma 
\Big(1+\alphanui (t)\Big)}
\left(\frac{\Enu}{\Enuiav\left(t\right)}\right)^{\alphanui(t)}\,
\nonumber\\
 & \,\,\,\, \times\,\, 
\exp\left[-\Big(1+\alphanui (t)\Big)\frac{\Enu}{\Enuiav(t)}\right]\,, 
\label{eq:phi}
\eeqa
where
\beqa
\alphanui (t)=\frac{2\Enuiav^2(t) - \Enuisqav(t)}{\Enuisqav (t) - 
\Enuiav^2 (t)}\nonumber
\eeqa 
is the spectral shape parameter. We will loosely refer to the 
quantity $F^0$ as the ``flux" with the understanding that the actual 
flux at Earth will be obtained by dividing the final emerging number 
of neutrinos from the SN per unit time and energy by $4\pi d^2$, $d$ 
being the distance from the Earth to the SN (assumed to be 10 kpc in the 
numerical calculations). For benchmark values of the quantities 
$L_{\nui}(t)$, 
$\Enuiav(t)$ and $\Enuisqav(t)$ we use those obtained from the results 
of the hydrodynamic simulations of the B/D   
group~\cite{Basel-Darmstadt-10} for a $18\msun$ progenitor star. These 
values are extracted for each time bin of the simulation and then time 
integrated to get the total flux. The variation of the luminosity and 
average energy over the emission time and the time-averaged 
normalized energy spectra of neutrinos of different flavors are 
shown in Figure \ref{fig:lum-avEnergy-timeprofile-spectra}. Note 
that all the $\nux$ and $\anux$ have the same luminosities, average 
energies and spectra in the B/D simulations. 

The neutrinos of different flavors emitted from their 
respective neutrinospheres can undergo a variety of flavor oscillation 
scenarios in passing through the SN matter. Unlike in  
many other neutrino sources, neutrinos and antineutrinos of 
all three flavors are present in core collapse SN explosions. This 
can give rise to interesting flavor oscillation scenarios. In 
addition, the extreme neutrino densities in the deep interior (first few 
hundred km) of SN can trigger collective neutrino 
oscillations due to neutrino-neutrino interactions; see, e.g., 
Refs.~\cite{Duan-etal-10, Mirizzi-etal-15} for reviews and references. 
While diffusing out of the exploding star these neutrinos will also go 
through the conventional Mikheyev-Smirnov-Wolfenstein (MSW) flavor 
oscillation at the outer layers of the SN matter far away from the 
self-induced (i.e., collective) oscillation zone. Because of the 
large spatial distance between the self-induced oscillation zone and 
the MSW oscillation zone, the MSW oscillation can happen essentially 
independently of the self-induced oscillations, with the latter 
simply pre-processing the neutrino flux of different flavors entering 
the MSW zone. 

The effect of these various oscillation processes on the final  
fluxes of neutrinos of different flavors emerging from the SN would of 
course depend on several factors like initial luminosity, average 
energies and flux hierarchy of different flavors, and as these 
properties change from one emission phase to another the oscillation 
scenarios also change. In the initial accretion phase, due to the 
extreme matter densities, the collective effects are generally 
considered to be matter suppressed, and only the usual MSW effect is 
present~\cite{Chakraborty-etal-11a, Chakraborty-etal-11b}. Although this 
remains to be conclusively established by further studies, we shall 
assume this to be the case in the present paper. Under this 
circumstance, the 
fluxes for the cases of normal ordering (NO) and inverted ordering (IO) 
of the neutrino mass hierarchy in the accretion phase are 
respectively given by (see, e.g., Ref.~\cite{Borriello-etal-12})
\begin{eqnarray}
F_{\anue}^{\NO} = \cos^2\theta_{12} F^0_{\anue} + 
\sin^2\theta_{12} F^0_{\anux}\quad {\mathrm{and}}\quad   
F_{\nu_e}^{\NO} = F^0_{\nu_x}\,, 
\label{eq:F_NO}
\end{eqnarray}
and 
\begin{eqnarray}
F_{\anue}^{\IO} = F^0_{\anux} \quad {\mathrm{and}}\quad  
F_{\nue}^{\IO} = \sin^2\theta_{12} F^0_{\nue} +
\cos^2\theta_{12} F^0_{\nux} \,\ .
\label{eq:F_IO}
\end{eqnarray}

In the longer cooling phase, due to smaller matter 
density, the dense matter suppression of collective effects is 
not present and multiple spectral splits due to self-induced 
conversions can happen in the collective 
oscillation region. This pre-processed flux will then go through the MSW 
oscillation. However, in the cooling phase the primary fluxes of 
different flavors of neutrinos are very similar (see 
Fig.~\ref{fig:lum-avEnergy-timeprofile-spectra}). Under this 
circumstance, the self-induced 
conversions would tend to equilibrate the fluxes among different 
flavors. Indeed, 
such a trend towards equilibration of the cooling 
phase neutrino fluxes of different flavors has been observed in a 
number 
of numerical studies; for detailed discussions see, e.g., 
\cite{Chakraborty-etal-11a,Borriello-etal-12,Fischer-etal-12}. 
Therefore, oscillation effects on the final cooling phase neutrino flux 
emerging from the SN may be expected to be minimal, and so the fluxes 
of neutrinos of different flavors emerging from the SN in the final 
cooling phase may also be taken to be essentially given by the same 
expressions as for the accretion phase given in equations (3) and (4) 
above. Indeed, since in this paper we are interested only in the 
time integrated flux, the small differences in the emerging fluxes of 
different neutrino flavors in each emission 
phase get smeared out. 

\section{The neutrino induced charged and neutral current cross sections and
the emission of neutrons from neutrino-excited nuclei}
\label{sec:xsecs-n-emission-etc}
The supernova neutrinos can excite the nuclei $^{A}{\X}_{N}$ of the 
detector material through the CC interaction producing the nuclei 
$^{A}{\Y}_{N-1}$ by the reaction
\begin{eqnarray}
\nue + ^A{\X}_N \to e^- + ^A{\Y}_{N-1}\,. 
\label{eq:nue-CC-general} 
\end{eqnarray}
The final nucleus gets de-excited by emitting particles or photons, and
we are interested here in the reactions where one or more neutrons are 
emitted: 
\begin{eqnarray}
^{A}{\Y}_{N-1} & \to & ^{A-1}{\Y}_{N-2} + n\,,
\label{eq:nue-CC-general-1n-emission}\\
^{A}{\Y}_{N-1} & \to & ^{A-2}{\Y}_{N-3} +2 n\,,
\label{eq:nue-CC-general-2n-emission}
\end{eqnarray}
and so on. For $\Fe$ detectors the excited nucleus is $\Co$ whereas
for $\Pb$ it is $\Bi$. In a similar fashion $\anue$ through
CC will produce positron and a different nucleus 
$^{A}{\Y}^{\prime}_{N+1}$ by the reaction 
\begin{eqnarray}
\anue + ^A{\X}_N \to e^+ + ^{A}{\Y}^{\prime}_{N+1}\,.
\label{eq:anue-CC-general}
\end{eqnarray}
The nucleus $^{A}{\Y}^{\prime}_{N+1}$ if excited above one or two 
emission threshold will emit neutrons in competition with other 
particles and photons. 

SN neutrinos can also excite detector nuclei through NC 
interactions. Neutrinos and antineutrinos of all three flavors can 
inelastically scatter from nuclei through NC interaction whereby the 
nucleus remains unchanged but goes to an excited state which can then 
de-excite by emitting neutrons. 

The CC neutrino capture or NC scattering process and the subsequent 
emission of neutrons can be considered as a two-step process, and like 
in Kolbe and Langanke~\cite{Kolbe-Langanke-01} and Engel 
et al~\cite{Engel-etal-03} we assume the two processes to be independent 
of each other. The first step involves calculating the CC or NC cross 
sections of neutrino interactions with the nuclei. During the 
interaction, an amount of energy $E_*\equiv E_f-E_i=\Enu-E_{\ell}$, 
where $E_i$ ($E_f$) is the energy of the initial (final)  
nucleus and $\Enu$ ($E_{\ell}$) is the energy of the incident 
neutrino (emerging lepton), is transferred to the final nucleus. This  
can excite a number of nuclear states in the final nucleus. The 
theoretical calculation of the cross section involves a multipole 
expansion of the current-current form of the weak interaction 
Hamiltonian and, assuming the lepton energies to be much greater than 
their masses, one relates the cross section to 
the nuclear transition operators of different multipole orders 
connecting the ground state of the target nucleus to the 
various (excited) states of the final nucleus 
\cite{Walecka-75,Kuramoto-etal-90,Kolbe-etal-03}. This gives the 
$\nu$-induced excitation spectrum, $\frac{d\sigma}{dE_*}(\Enu,E_*)$, of 
the final nucleus, i.e., the differential cross section as a function 
of the excitation energy $E_*$ of the final nucleus. 
We then calculate, in the second step, 
the cross section for emission of one-, two or more neutrons by the 
final nucleus by folding the $\nu$-induced excitation spectrum of the 
final nucleus with the probabilities of decay of the nucleus through 
emission of specified number of neutrons as a function of the 
excitation energy of the nucleus. The latter is calculated 
using a suitable nuclear statistical model code discussed below.  

For $\nue$ CC interactions, at typical SN neutrino energies of few tens 
of MeV for which $q=|\boldq| \to 0$ limit ($\boldq$ being the 3-momentum 
transfer to the nucleus) is applicable, the cross section is dominated 
by contributions from the allowed transitions to the isobaric analog 
state (IAS) and the Gamow-Teller (GT) resonance states in the final 
nucleus. The allowed contribution to the $(\nue,e^-)$ differential 
cross section in the 
$q\to 0$ limit can thus be written in terms of the effective Fermi and GT 
transition strengths 
as~\cite{Kuramoto-etal-90,Fuller-etal-99,Kolbe-etal-99} 
\beqa
\frac{d\sigma_{\nue}^{\rm CC, allowed}}{dE_*}(\Enu,E_*) = 
\frac{G_F^2 \cos^2\theta_c}{\pi} \,\,\, p_e E_e F(Z+1,E_e)\nonumber\\
  \hskip-1cm \times \left(S_{\F}(E_*) +   
S_{{\GT}_{-}}(E_*)\right),
\label{eq:diff_sigma_CC}
\eeqa
where $m_e$, $p_e$ and $E_e$ are the mass, momentum and energy of the 
emitted electron, respectively, $G_F$ is the Fermi constant, $\theta_C$ 
is the Cabibbo angle, 
\beq
S_{\F}(E_*)=\frac{1}{2J_i+1}\big\vert \langle 
J_f\|\sum_{k=1}^{A}\tauplus(k)\|J_i\rangle\big\vert^2
\label{eq:S_F}
\eeq
and 
\beq
S_{{\GT}_{-}}(E_*)=\frac{(g_A^{\rm eff})^2}{2J_i+1}\big\vert \langle 
J_f\|\sum_{k=1}^{A}\tauplus(k) {\bg\sigma} 
(k)\|J_i\rangle\big\vert^2
\label{eq:S_GT}
\eeq
are the Fermi and GT strength distributions, respectively, as functions 
of the excitation energy $E_*=E_f-E_i = E_{\nu}-E_e$ of the final 
nucleus, $J_i$ and $J_f$ represent the angular momentum of the initial 
and final nucleus, respectively, $\tauplus$ is the operator that 
converts a neutron to a proton, and $\bg\sigma$ are the standard Pauli 
spin matrices. The quantity $(g_A^{\rm eff})$ is the ratio of the 
effective axial vector to vector coupling constants of the nucleon in 
the $q\to 0$ limit, whose value for the ``bare" nucleon is 
1.26~\cite{Kuramoto-etal-90,Fuller-etal-99}. The 
factor $F(Z+1,E_e)$ is the correction factor to account for the 
distortion of the outgoing electron wave function due to the Coulomb 
field of the 
final state nucleus. For $p_{e,{\rm eff}} R \ll 1$, where $p_{e,{\rm 
eff}} = \left[\Big(E_e - V(0)\Big)^2 - m_e^2\right]^{1/2}$ 
with 
$V(0)=(3/2)e^2 Z_f/R$, ($e Z_f$ and $R$ being the charge and radius of 
the final nucleus, respectively), the Coulomb correction factor can be 
represented by the Fermi function~\cite{Engel-98}, 
\beq
F(Z,E)=2 (1+\gammazero)(2p_e R)^{2(\gammazero-1)} 
\frac{\arrowvert\Gamma(\gammazero+iy)\arrowvert^2}{ 
\arrowvert\Gamma(2\gammazero+1)\arrowvert^2}\exp(\pi y)\,,
\label{eq:FermiFunction}
\eeq
where $\gammazero=(1-Z^2\alpha^2)^{1/2}$ and $y=\alpha Z E_e/p_e$, with 
$\alpha$ the fine structure constant.  

The $\anue$ CC cross section is similar where the final state 
lepton is a positron instead of electron, and the Fermi function then 
takes care of the distortion of the positron wave function due to the 
repulsion effect of the final nucleus. Also, one replaces the operator 
$\tauplus$ above by $\tauminus$ which transforms a proton to a neutron. 

For the NC scattering of neutrino, the analog of Fermi transition only 
contributes to elastic scattering~\cite{Fuller-etal-99}, and the allowed 
contribution to the cross section for inelastic scattering (that 
involves energy transfer to the target nucleus) in the $q\to 0$ 
limit is governed by the NC GT (the so-called ${\GT}_0$) 
strength~\cite{Fuller-etal-99}
\beq
S_{{\GT}_{0}}(E_*)=\frac{(g_A^{\rm eff})^2}{2J_i+1}\big\vert \langle
J_f\|\sum_{k=1}^{A}\half\tau_3 (k) {\bg\sigma}
(k)\|J_i\rangle\big\vert^2\,,
\label{eq:S_GT0}
\eeq
giving the allowed contribution to the $(\nu,\nu^{\prime})$ NC induced
excitation spectrum of the final nucleus in the
$q\to 0$ limit as 
\beq
\frac{d\sigma_{\nu}^{\rm NC, allowed}}{dE_*}(\Enu,E_*) =
\frac{G_F^2}{\pi} \,\, E_{\nu^{\prime}}^2
S_{{\GT}_{0}}(E_*)\,,
\label{eq:diff_sigma_NC}
\eeq
where $E_{\nu^{\prime}}=\Enu - E_*$ is the energy of the final neutrino. 
In equation (\ref{eq:S_GT0}), $\half\tau_3$ is the third (``z"-) 
component of the isospin operator. 
 
For small but non-zero momentum transfer $q$, in addition to the allowed 
contributions to the cross sections discussed above,  
there would be additional small, but in general non-negligible, 
contributions to the  
cross sections due to the forbidden transitions originating from terms 
linear in $q$~\cite{Kuramoto-etal-90}. 

In the case of $\Fe(\nue,e^-)\Co$ 
reaction, with the ground state of 
$\Fe$ being a $J=0$ state, the allowed transitions connect to the $0^+$ 
and $1^+$ states of $\Co$. We add to that the smaller contributions 
coming from the first forbidden transitions to the $1^-$ and $2^-$ 
states in $\Co$ which are mostly at higher excitation energies than the 
ones connected by the allowed transitions. For the allowed transitions, 
we use the GT strength distribution from large-scale shell model 
calculations using the monopole corrected KB3 interaction 
\cite{Caurier-etal-99} which is very successful in 
reproducing the spectra of nuclei in the iron region. The calculated GT 
strengths ${\GT}_{-}$ and ${\GT}_{+}$ for the $\Fe(\nu_e,e^{-})\Co$ and 
$\Fe(\bar\nu_e,e^{+})\Mn$ reactions obtained by Caurier \etal 
\cite{Caurier-etal-99} compare well with the values extracted from $L=0$ 
forward angle $(p,n)$ and $(n,p)$ cross sections, respectively, with 
some small differences appearing at the high energy end. The 
strength distributions for the forbidden transitions to $1^-$ and $2^-$ 
states of $\Co$ are obtained by unfolding the contributions of 
these transitions to the neutrino-spectrum-averaged $\Fe(\nue,e^-)\Co$ 
differential cross section as a function of the $\Co$ excitation energy 
given in Ref.~\cite{Kolbe-etal-99} for the KARMEN experiment's 
decay-at-rest $\nue$ spectrum~\cite{KARMEN-Maschuw-98}.  

For the NC case, we use the strength distribution calculated by Toivanen 
\etal \cite{Toivanen-etal-01} which includes both the
$\Delta T$=0 and 1 resonances though the strength for transitions to
the higher isospin is very small.

With the $\nu$-induced excitation spectrum of the final nucleus obtained 
as above, the total CC or NC cross section as a function of the 
incident neutrino energy $\Enu$ is obtained as 
\beq
\sigma^{\rm CC (NC)} (\Enu)=\int 
\frac{d\sigma^{\rm CC (NC)}}{dE_*}(\Enu,E_*) dE_*\,,
\label{eq:total-xsec}
\eeq
and the CC and NC cross sections for emission of one-, two- or 
three neutrons as 
\beq
\sigma_{1n (2n) (3n)}^{\rm CC (NC)} (\Enu)=\int 
\frac{d\sigma^{\rm CC (NC)}}{dE_*}(\Enu,E_*) P_{1n (2n) 
(3n)}(E_*)dE_*\,, 
\label{eq:n-emission-xsec}
\eeq
where $\frac{d\sigma^{\rm CC (NC)}}{dE_*}(\Enu,E_*)$ is the relevant 
$\nu$-induced excitation spectrum of the final nucleus, 
and $P_{1n}(E_*)$, $P_{2n}(E_*)$, $P_{3n}(E_*)$ are the probabilities 
for emission of one-, two- and three neutrons, respectively, as 
functions of the excitation energy of the final nucleus. 

\begin{figure*}[!]
\epsfig{file=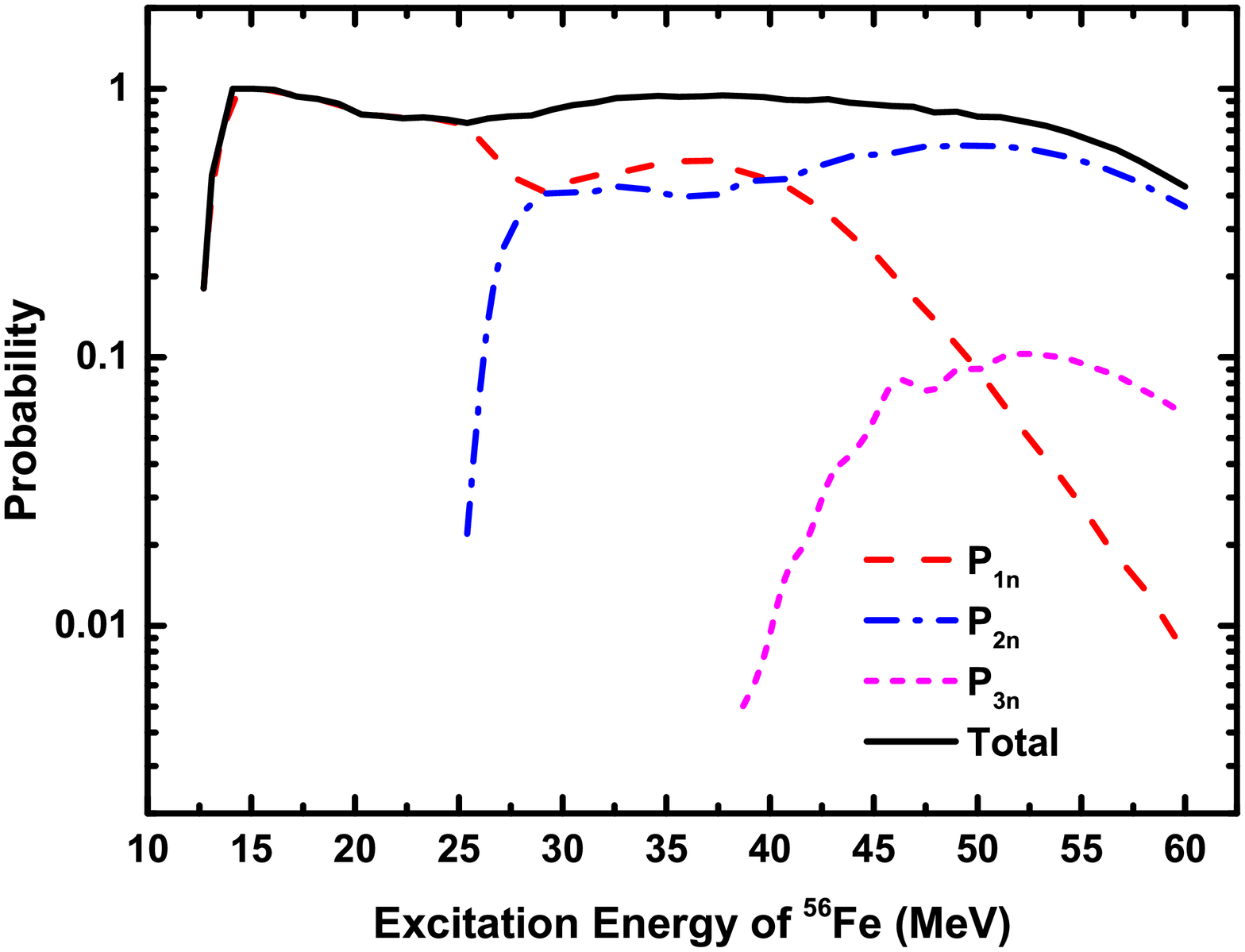,width=\columnwidth}
\epsfig{file=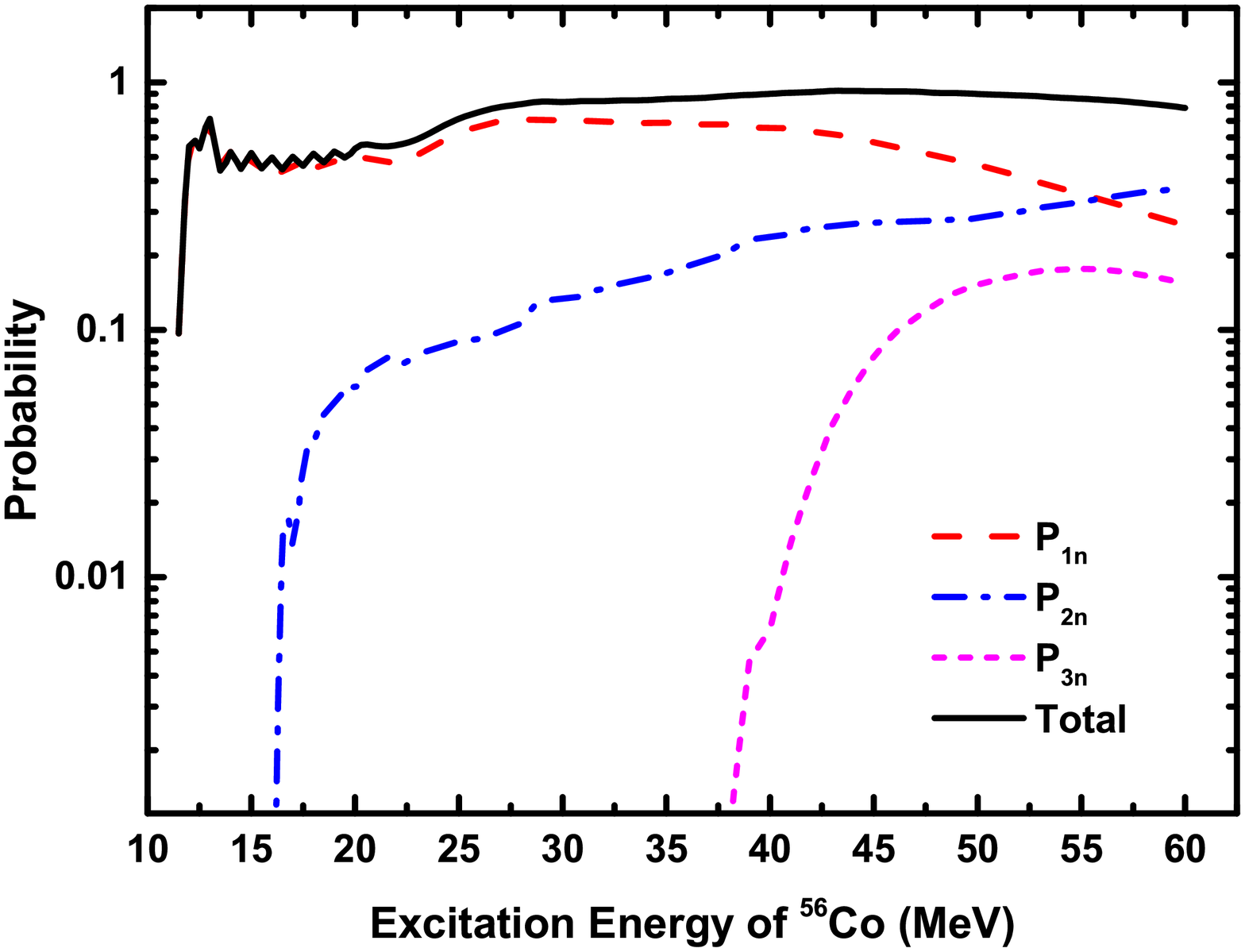,width=\columnwidth}
\caption{The one-, two- and three neutron emission probabilities 
($P_{1n}$, $P_{2n}$, and $P_{3n}$, respectively) of excited $\Fe$ 
nucleus (left) and excited $^{56}{\rm Co}$ nucleus 
(right) as a function of the excitation energy.} 
\label{fig:n-emission-prob-exc-56Fe-56Co}
\end{figure*}

Finally, the number of neutrons produced as a function of incident 
neutrino energy is given by  
\beqa
\frac{dN_n^{\rm CC (NC)}}{d\Enu}(\Enu) = N_T \Phi_{\nu}(\Enu) 
\sum_{k=1}^3 k\sigma_{kn}^{\rm CC (NC)}(\Enu)\,,
\label{eq:n-number}
\eeqa
where $\sigma_{kn}^{\rm CC (NC)}$ (for $k=1,2,3$) are given by equation 
(\ref{eq:n-emission-xsec}), $\Phi_{\nu}(\Enu)$ is the time-integrated flux 
spectrum (number per unit area per unit energy) of the SN neutrinos 
falling on the detector, and $N_T$ is the total number of target 
detector nuclei. 

The calculations of the neutron emission probabilities for $\Fe$ were 
done using 
the fusion-evaporation code PACE4~\cite{PACE4-code} originally developed 
by Gavron~\cite{Gavron-80}, which is based on a modified version of 
Monte Carlo statistical model calculations using angular momentum 
projection at each stage of de-excitation of the nucleus. 
The results for the neutron emission probabilities from 
excited $\Fe$ and $^{56}{\rm Co}$ nuclei as functions of excitation 
energy are shown in Figure~\ref{fig:n-emission-prob-exc-56Fe-56Co}. 
These are used for the calculations of
emission of neutrons through NC interaction of neutrinos of all flavors
and CC interaction of $\nu_e$, respectively, with $^{56}{\rm Fe}$ 
nuclei. Note that the excitation energy threshold for emission of one  
neutron (1n) from $\Fe$ is 11.2 MeV whereas it is 25 MeV for two 
neutrons (2n). 
Three neutrons are emitted only beyond the high excitation energy of
38 MeV. In the case of excited $^{56}{\rm Co}$ nuclei resulting from CC 
interaction of $\nue$ with $\Fe$, the 1n emission starts from 
excitation energy of 11 MeV and 2n from 16 MeV. Three neutrons 
are emitted only beyond an excitation energy of 35 MeV. 
Results for CC interaction of $\anue$ with $\Fe$ resulting in 
excited $^{56}{\rm Mn}$ are also treated in a similar manner.

For the case of $^{208}{\rm Pb}$ we directly use the 1n and 2n emission 
cross sections given by Engel, McLaughlin and Volpe 
\cite{Engel-etal-03} which have the weak interaction cross section 
folded in for both CC and NC. These calculations were carried out by the 
coordinate space Skyrme-Hartree-Fock method in a 20 fm box and then by a 
version of RPA using the same Skyrme interaction in the basis of 
Hartree-Fock states. The threshold for 1n and 2n emission for 
$^{208}{\rm Pb}$ are 6.9 and 15.0 MeV, respectively.

%
\section{Results}
\label{sec:results}
\subsection*{Pb detectors}
\begin{figure*}[!]
\epsfig{file=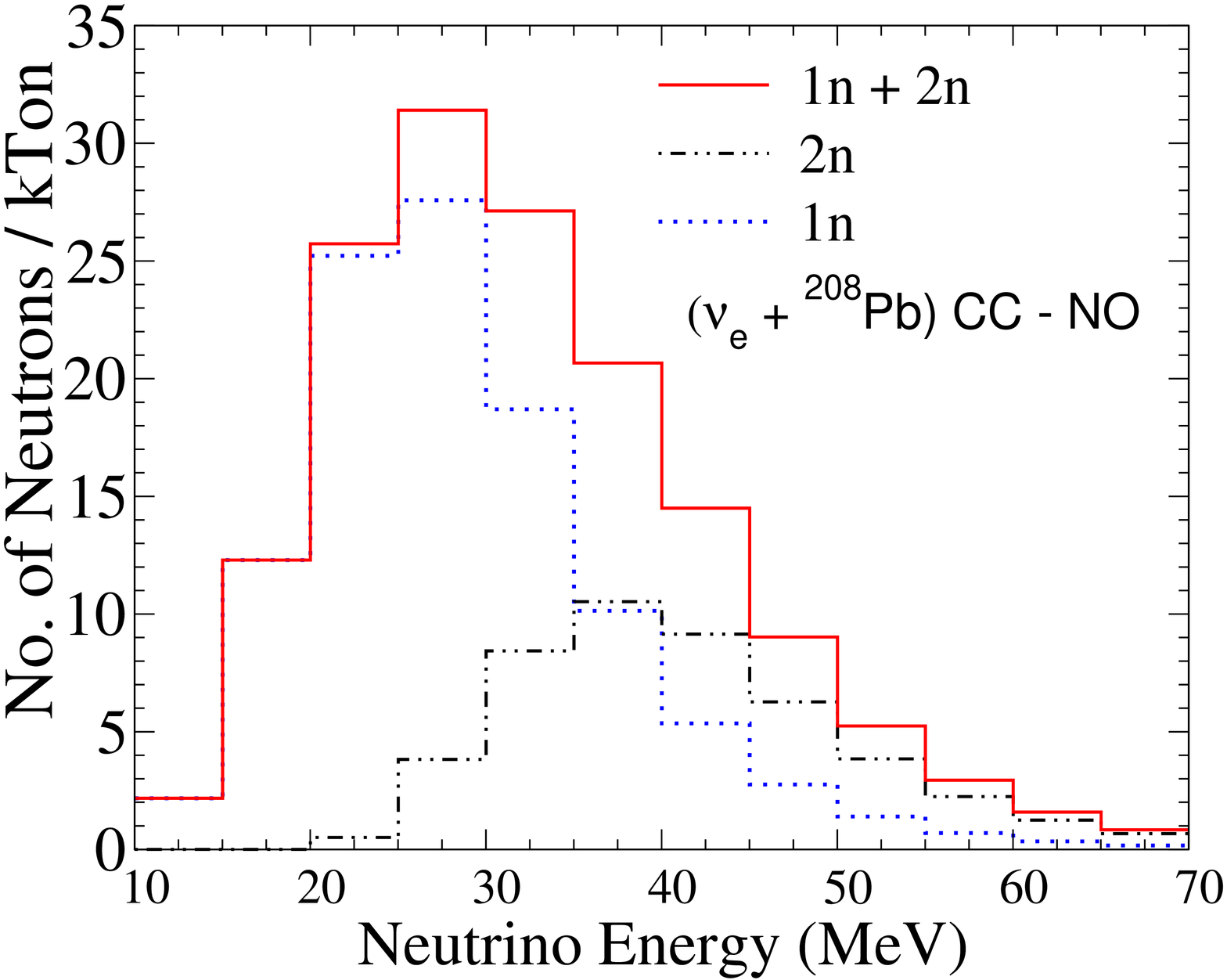,angle=270,width=\columnwidth}
\epsfig{file=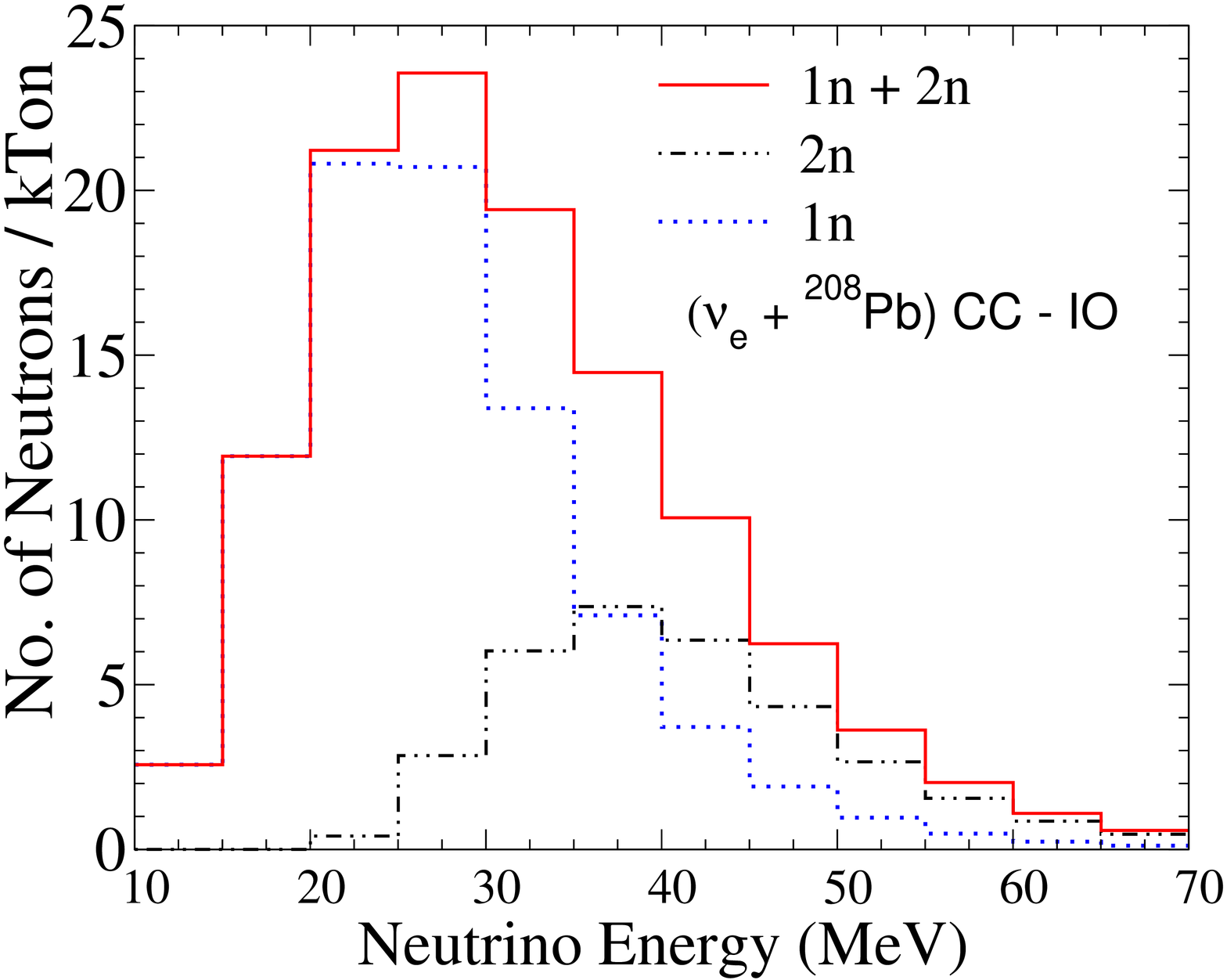,angle=270,width=\columnwidth}
\caption{Number of neutrons emitted due to $\nu_e$ CC
interactions per kTon of $\Pb$ as function of neutrino energy for  
Normal Order (NO) (left panel) and Inverted Order (IO) (right panel) of 
neutrino mass hierarchy. ``1n" and ``2n" indicate neutrons from one- and 
two-neutron emission events, respectively.} 
\label{fig:neutrons-nue-CC-Pb_a:NO_b:IO}
\end{figure*}

The number of neutrons emitted per kTon of lead for CC interaction of 
supernova $\nue$'s as a function of the neutrino energy for   
the time-integrated neutrino fluxes given by the B/D simulation 
results discussed above in Section \ref{sec:ccSN-nu} is given in the 
form of a histogram of bin size 5 MeV in Figure 
\ref{fig:neutrons-nue-CC-Pb_a:NO_b:IO} for both Normal Order (NO) 
and Inverted Order (IO) of neutrino mass hierarchy assuming 100\% 
detection efficiency. The bin-wise contributions of 1n and 2n events are 
also shown. Contributions from 3n events are negligible.  
\begin{figure}
\includegraphics[width=0.8\columnwidth,angle=270]{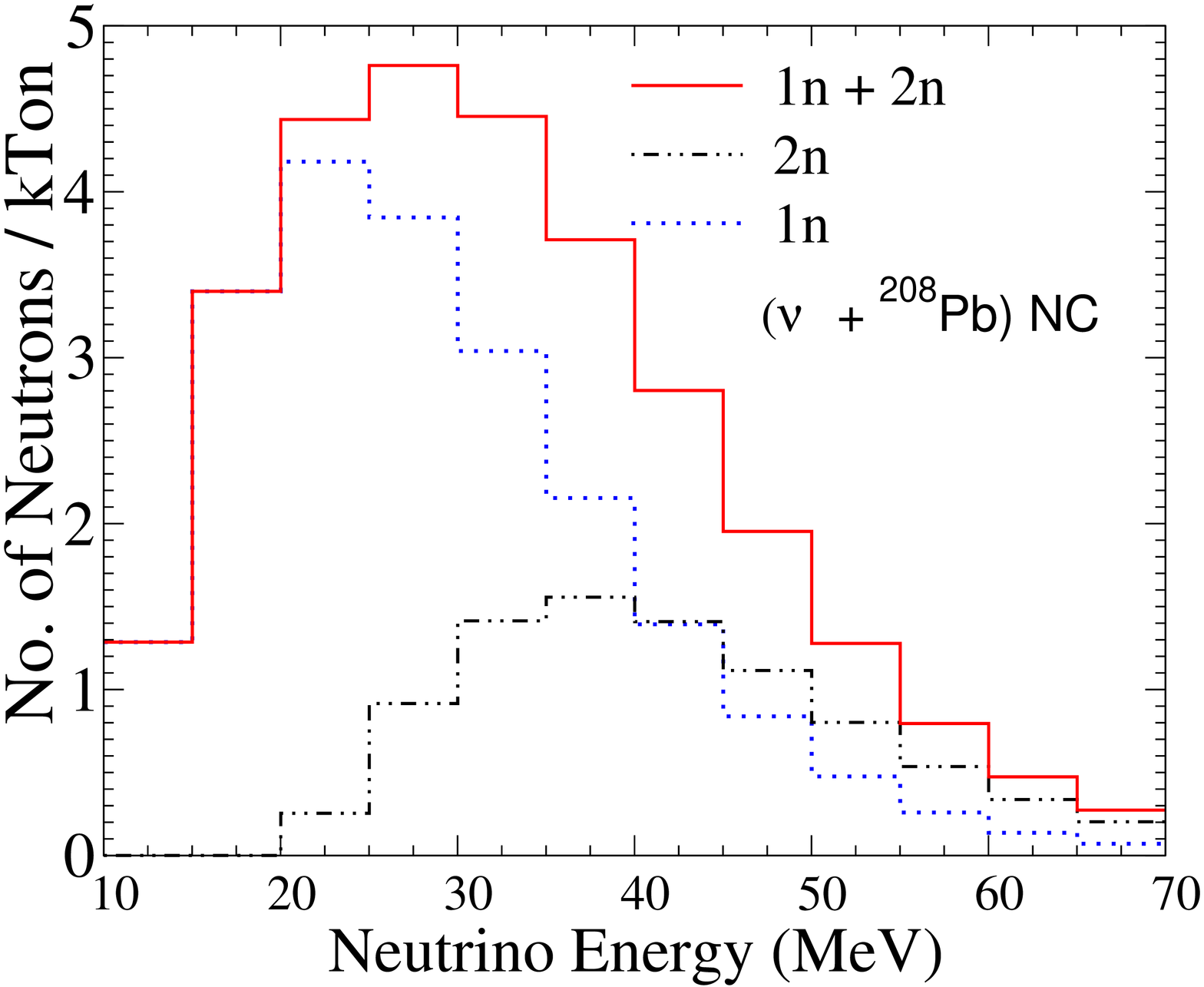}
\caption{Number of neutrons emitted due to NC interaction 
of all six species of neutrinos per kTon of $\Pb$ as function of 
neutrino energy. ``1n" and ``2n" indicate neutrons from one- and 
two-neutron emission events, respectively.}
\label{fig:neutrons-NCtotal-Pb}
\end{figure}
These take into account the matter effect as described in Sec.\ 
\ref{sec:ccSN-nu}. From the right panel of 
Fig.~\ref{fig:lum-avEnergy-timeprofile-spectra}, which shows the 
time-integrated energy spectra for all three neutrino species, we see 
that the maximum in the energy distribution of the electron 
neutrinos is below 10 MeV beyond which the spectrum falls rapidly. 
However, the CC cross section increases rapidly with neutrino energy, 
roughly as $\Enu^2$. In addition, the probability of one neutron 
emission from excited $\Bi$ nucleus starts increasing and taking 
non-zero values beyond the 1n emission threshold of 6.9 MeV. 
Together, these effects result in a maximum for the number of neutrons 
emitted in the energy bin 
of 25-30 MeV for the $\nu_e$ case both for NO and IO of the mass 
hierarchy. The total number of neutrons emitted in the 
case of NO is 154, with 1n and 2n events contributing 107 and 47 
neutrons, respectively. The corresponding numbers for the case of IO 
are 117, 84 and 33, respectively. 

We note here that the above numbers are significantly smaller than 
earlier estimates (see, e.g., \cite{Scholberg:2012id}) based on older SN 
models \cite{Totani-etal-98,Gava-etal-09} which have different neutrino 
energy spectra compared to those given by the B/D simulations considered 
here and also involve different neutrino flavor conversion scenarios. 
One observes that for 
$\nu_e$ the average energy in the B/D simulation is always 
less than 15 MeV and below 10 MeV for the longest lasting cooling phase 
(see Fig.~\ref{fig:lum-avEnergy-timeprofile-spectra}). 
Similarly, the $\nu_x$ average energies are always lower than 17 MeV and
lower than 14 MeV during the entire cooling phase. With both $\nu_e$ and
$\nu_x$ having lower average energies than those in earlier simulations, 
the number of neutrons for both NO and IO are smaller than the 
earlier estimates (even with 100\% detection efficiency and 
substantial flavor mixing assumed here), and so is the difference 
between the numbers for NO and IO cases. 

We also mention here that, 
since a detector such as the HALO \cite{HALO_detector} may be able to 
distinguish between the 1n and 2n events, the ratio of 1n and 2n events 
(which would be independent of the total neutrino flux) may yield 
interesting spectral information on the SN neutrino 
flux~\cite{Vaananen-Volpe-11}. This, however, is subject to several 
uncertainties, and we will not discuss this further in this paper. 

The number of events due to CC interactions of the SN $\anue$'s 
($^{208}{\rm Pb} + \anue \to ^{208}{\rm Tl} + e^+$) are orders of
magnitude smaller than those due to $\nue$'s mainly due to the fact that 
beta transitions of protons to neutrons are strongly suppressed due to 
Pauli blocking of the neutron single particle states. Thus the process 
has negligibly small contribution and we do not consider the $\anue$ 
events. 

For the NC events all the six species of supernova neutrinos contribute 
and there is no difference in the results for the two mass hierarchies. 
The total number of neutrons emitted for NC interactions per kTon of 
lead is 30, with 21 from 1n events and the rest from 2n events. The 
histogram of the number of neutrons emitted as a function of neutrino 
energy with a bin size of 5 MeV is shown in 
Fig.~\ref{fig:neutrons-NCtotal-Pb}. These numbers are also smaller than 
the previous estimates~\cite{Scholberg:2012id}. 

It is clear that the number of neutrons produced per kTon of $\Pb$ is 
dominated by the CC events, a result already known from 
earlier studies; see, e.g., Ref.~\cite{Vaananen-Volpe-11}. From 
the numbers obtained above, the number of neutrons produced through NC 
interactions is $\sim$ 16\% (20\%) of the total number of produced 
neutrons in the case of NO (IO) of mass hierarchy. Thus, in absence
of the capability to identify the CC induced events (through
identification of the accompanying $e^-$) it will not be possible for a 
$\Pb$ detector by itself to determine the fraction of the $\mu$ 
and $\tau$ flavored neutrinos in the total SN neutrino flux. 
\subsection*{Fe detectors}
The total CC cross section ($\sigma^{\rm\sss CC}$) for
$\Fe(\nu_e,e^-)\Co$ and total NC cross section
($\sigma^{\rm\sss NC}$) for
$\Fe(\nu,\nu^{\prime})\Fe$ as functions of neutrino 
energy ($E_\nu$), with the GT strength distributions as described in 
Sec.~\ref{sec:xsecs-n-emission-etc} including the contributions 
coming from the first forbidden transitions, are given in 
Table \ref{tab:cc1}. The corresponding 
one- and two neutron emission cross sections, 
$\sigma_{1n}^{\rm CC (NC)}$ and $\sigma_{2n}^{\rm CC (NC)}$, are also 
tabulated. 
\begin{table}
{\footnotesize 
\begin{tabular}{|c|c|c|c||c|c|c|}
\hline
\multicolumn{1}{|c|}{} &  
\multicolumn{3}{|c|| }{$\Fe (\nu_e,e^-) \Co$} & 
\multicolumn{3}{c| }{$\Fe (\nu,\nu^\prime) \Fe$}\\ 
\hline 
$\Enu$ & $\sigma^{\rm\sss CC}$ & $\sigma_{1n}^{\rm\sss CC}$ & 
$\sigma_{2n}^{\rm\sss CC}$ & $\sigma^{\rm\sss NC}$ & 
$\sigma_{1n}^{\rm\sss NC}$ & $\sigma_{2n}^{\rm\sss NC}$\\
(MeV) & (${\cm}^2$) & (${\cm}^2$) & (${\cm}^2$) & (${\cm}^2$) & 
(${\cm}^2$) & (${\cm}^2$) \\     
\hline
  10   & 3.12E-42    & 0.0       & 0.0         
       & 1.37E-43    & 1.31E-44  & 0.0 \\
  15   & 1.31E-41   & 2.89E-43   & 0.0         
       & 2.23E-42   & 2.80E-43   & 0.0\\
  20   & 3.65E-41   & 2.50E-42   & 9.33E-46    
       & 7.86E-42   & 1.21E-42   & 0.0   \\
  25   & 7.43E-41   & 7.22E-42   & 1.26E-44    
       & 1.73E-41   & 2.99E-42   & 1.16E-49 \\
  30   & 1.28E-40   & 1.50E-41   & 9.65E-44    
       & 3.05E-41   & 5.70E-42   & 2.77E-45 \\
  35   & 2.02E-40   & 2.80E-41   & 4.83E-43    
       & 4.74E-41   & 9.42E-42   & 3.68E-44 \\
  40   & 3.04E-40   & 4.94E-41   & 1.60E-42    
       & 6.82E-41   & 1.42E-41   & 1.49E-43\\
  45   & 4.45E-40   & 8.45E-41   & 4.08E-42    
       & 9.27E-41   & 2.01E-41   & 3.88E-43\\
  50   & 6.39E-40   & 1.40E-40   & 8.72E-42    
       & 1.21E-40   & 2.71E-41   & 8.13E-43 \\
  55   & 9.03E-40   & 2.23E-40   & 1.66E-41    
       & 1.53E-40   & 3.53E-41   & 1.49E-42\\
  60   & 1.26E-39   & 3.43E-40   & 2.88E-41    
       & 1.89E-40   & 4.47E-41   & 2.49E-42\\
\hline
\end{tabular}
}
\caption{The total CC cross section ($\sigma^{\rm\sss CC}$) for
$\Fe(\nu_e,e^-)\Co$ and total NC cross section
($\sigma^{\rm\sss NC}$) for $\Fe(\nu,\nu^{\prime})\Fe$ as functions of 
neutrino energy ($E_\nu$). The corresponding one- and two neutron 
emission cross sections, $\sigma_{1n}^{\rm CC (NC)}$ and 
$\sigma_{2n}^{\rm CC (NC)}$, are also given.}
\label{tab:cc1}
\end{table}

\begin{figure}
\includegraphics[width=0.85\columnwidth]{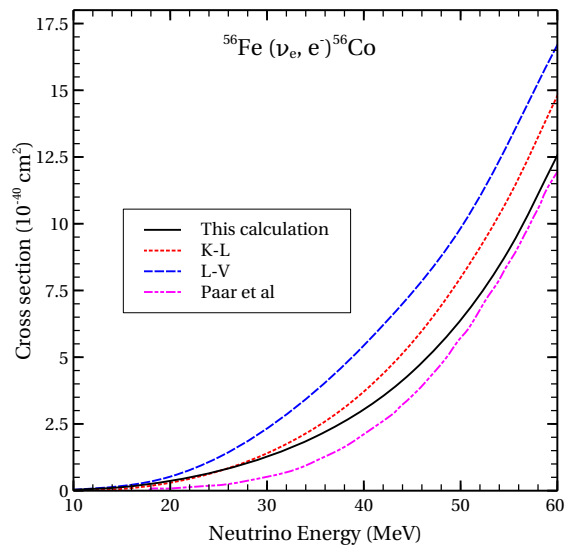}
\caption{Total $\Fe(\nu_e,e^-)\Co$ cross section as a function of 
neutrino energy. For comparison, the cross sections 
obtained in Refs.~\cite{Kolbe-Langanke-01} (K-L), 
\cite{Lazauskas-Volpe-07} (L-V) and \cite{Paar-etal-08} (Paar et al) are 
also shown.}  
\label{fig:nue-56Fe_CCxsec-comparison}
\end{figure}

The total $\Fe(\nu_e,e^-)\Co$ CC cross section has been calculated by 
several authors using various microscopic approaches. In Figure 
\ref{fig:nue-56Fe_CCxsec-comparison} we show a comparison of our results 
with those given in 
Refs.~\cite{Kolbe-Langanke-01,Lazauskas-Volpe-07,Paar-etal-08}, for 
example. It is seen that, while all the results show a rapidly rising 
cross section with neutrino energy, there is considerable 
amount of differences amongst the results of different theoretical 
calculations. Our results are close to those of Kolbe and Langanke 
\cite{Kolbe-Langanke-01} (K-L) at lower energies while being close to 
the results of \cite{Paar-etal-08}(Paar et al) at relatively higher 
energies. At all energies, the cross section values obtained in 
\cite{Lazauskas-Volpe-07} (L-V) are the largest, while those given by  
Paar et al \cite{Paar-etal-08} are the smallest, the two sets of values 
differing by almost a factor of 5 at some energies. Our results lie 
bracketed within those of K-L and Paar et al at all energies. 

The $\Fe(\nu_e,e^-)\Co$ cross section has been 
measured, although with a low precision, by the KARMEN 
experiment~\cite{KARMEN-Maschuw-98}. The spectrum-averaged 
value of the cross section for the 
decay-at-rest (DAR) $\nue$ spectrum, $n(\Enu)_{\rm DAR}=96 \Enu^2 (M_\mu 
- 2\Enu)/M_\mu^4$ (where $M_\mu$ is the muon mass) was measured to be 
$\langle\sigma(\Fe(\nu_e,e^-)\Co)\rangle_{\rm DAR} = 
[2.51 \pm 0.83 ({\rm stat.}) \pm 0.42 ({\rm syst.})]\times 10^{-40}
\cm^2$. With our $\nue$ CC cross section values 
given in Table \ref{tab:cc1}, we get a value of 
$1.98\times10^{-40}\cm^2$ for the same DAR spectrum-averaged cross 
section, which is quite consistent within experimental errors with the 
KARMEN measured value.   

Our results for the 
total NC cross section for the process $\Fe(\nu,\nu^{\prime})\Fe$ given 
in Table \ref{tab:cc1} also compare reasonably well with those given in 
Table VI of K-L~\cite{Kolbe-Langanke-01}, for example.  

\begin{figure*}[!]
\epsfig{file=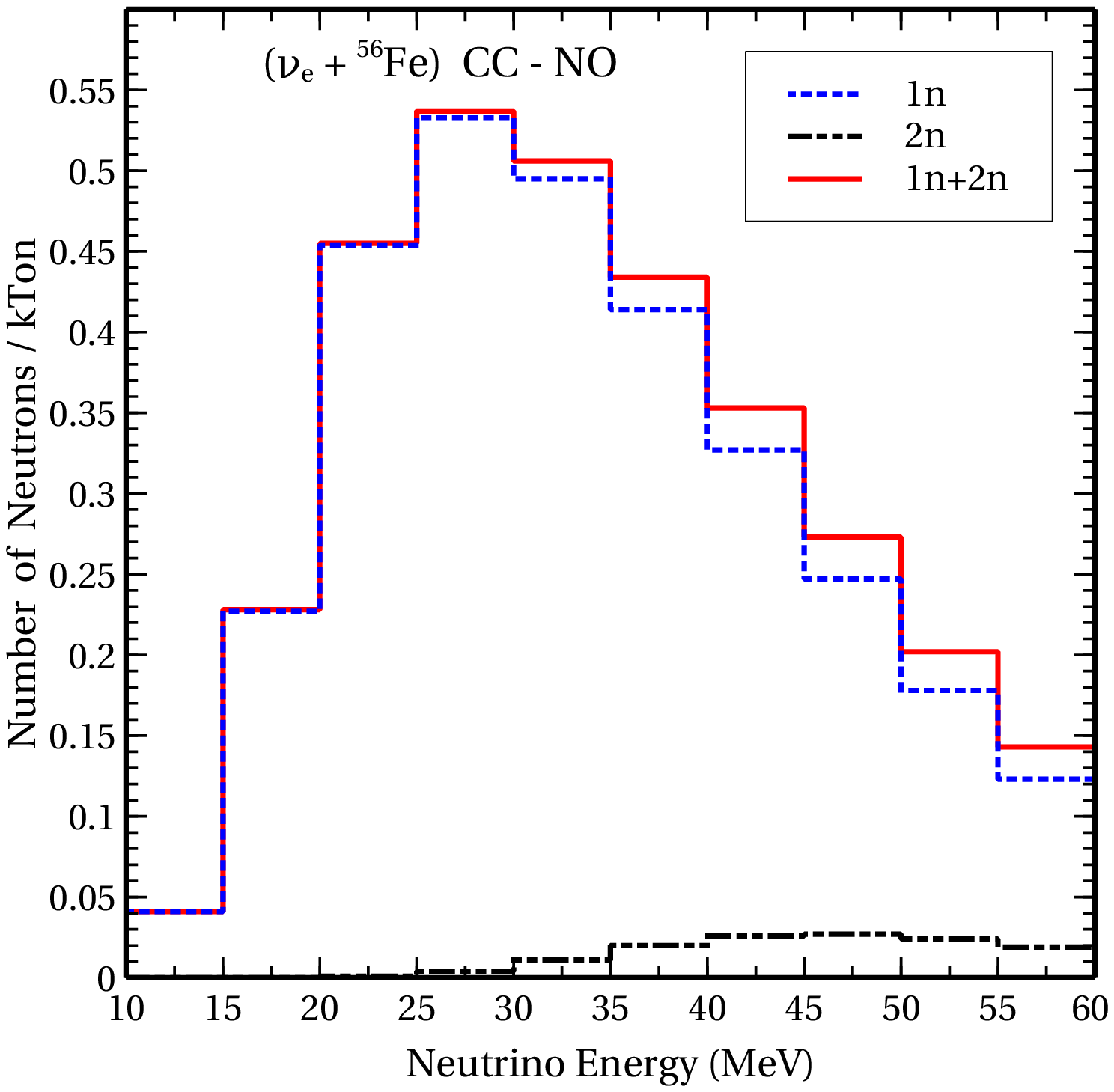,width=0.4\hsize}\hskip 
0.5cm
\epsfig{file=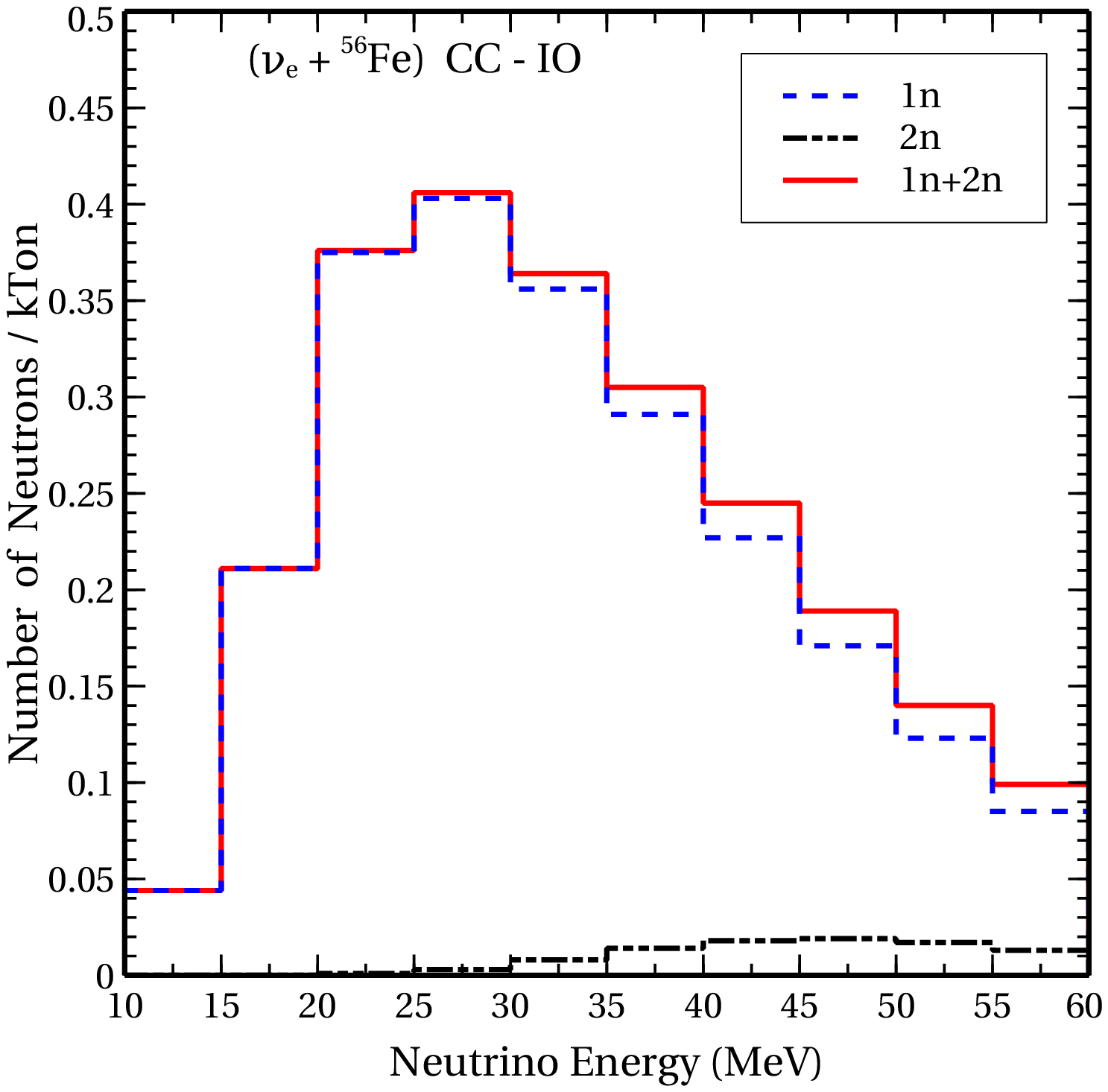,width=0.4\hsize}
\caption{Number of neutrons emitted due to $\nu_e$ CC
interactions per kTon of $\Fe$ as function of neutrino energy for  
Normal Order (NO) (left panel) and Inverted Order (IO) (right panel) of 
neutrino mass hierarchy. ``1n" and ``2n" indicate neutrons from 
one- and two-neutron emission events, respectively.}
\label{fig:neutrons-nue-CC-Fe_a:NO_b:IO}
\end{figure*}

The number of neutrons emitted per kTon of $\Fe$ as a function of 
incoming neutrino energy, calculated using 
equation (\ref{eq:n-number}) with the $\Fe (\nue, e^-)\Co$ CC one- and 
two neutron emission cross sections given in Table \ref{tab:cc1}, is 
shown in Fig.~\ref{fig:neutrons-nue-CC-Fe_a:NO_b:IO} in the form 
of histograms of 5 MeV energy bins for both NO and IO mass hierarchies. 
It is seen that, unlike in the case of $\Pb$, essentially all the 
neutrons are from 1n events, with 2n events giving negligibly small   
contribution in the case of $\Fe$. 
The total number of neutrons emitted 
is $\sim$ 3 and 2 for NO and IO, 
respectively. As the neutron emission threshold for 
$\Co$ is $\sim$ 11.0 MeV, only the tail of the neutrino
energy distribution contributes. Also the Fermi strength at the Isobaric
Analog State (IAS) around 4 MeV and most of the GT giant 
resonance distribution has no contribution as the one neutron emission 
probability
is zero or almost zero up to $\sim$ 12.5 MeV. As a result the forbidden 
transitions become relatively more important ($\sim$ 24\% and 22\%
of the GT strengths for NO and IO cases, respectively). However, even 
including those, the total number of neutron events turns out to be very 
small.

The $\anue$ CC process, $^{56}{\rm Fe}(\anue,e^+)^{56}{\rm Mn}$, 
gives negligibly small number of neutrons emitted by the excited 
$^{56}{\rm Mn}$ nucleus. Firstly, the total ${\GT}_+$ strength is 
experimentally seen to be $\sim$ 2.8 with the theoretical shell model 
number being 2.7~\cite{Caurier-etal-99}, whereas the total ${\GT}_-$ 
strength corresponding to the $\nue$ CC reaction 
$^{56}{\rm Fe}(\nu_e,e^-)^{56}{\rm Co}$ discussed above is 9.9 
experimentally~\cite{Caurier-etal-99} and 9.3 from theoretical 
calculations. But even more important is the fact that the nucleus 
$^{56}{\rm Mn}$ has the one-neutron emission threshold
above 8 MeV and almost all the ${\GT}_+$ strength lies below the 
excitation energy of 8 MeV. 

\begin{figure}
\includegraphics[width=0.9\columnwidth]{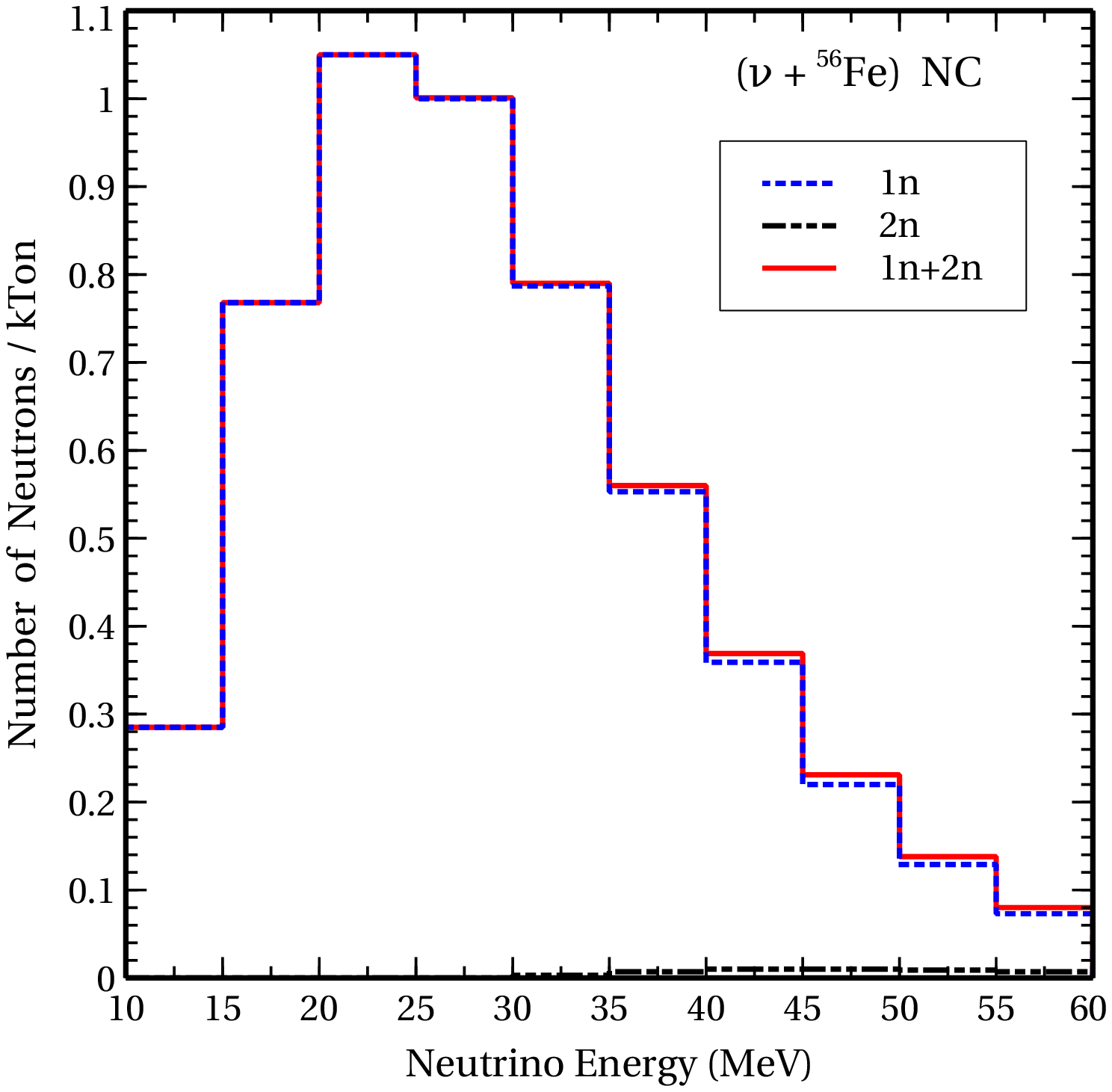}
\caption{Number of neutrons emitted due to NC 
interaction of all six species of neutrinos per kTon of $\Fe$ as function 
of neutrino energy. ``1n" and ``2n" indicate neutrons from one- 
and 
two-neutron emission events, respectively.}
\label{fig:neutrons-NCtotal-Fe}
\end{figure}

For NC interactions, the histogram of the sum of the number of neutrons 
emitted due to all the six species of neutrinos is shown in 
Fig.~\ref{fig:neutrons-NCtotal-Fe}. Again, the contribution 
from 2n events is negligibly small. The total number of neutrons emitted 
is $\sim$ 5. Without the additional contribution from $(\Delta T=1)$ 
resonance, the number is $\sim$ 4. This again is the reflection of 
the fact that the 1n emission threshold for $^{56}{\rm Fe}$ is 
8 MeV whereas the NC excitation strength (${\GT}_0$) distribution has 
only the tail part of the resonance beyond 8 MeV 
\cite{Toivanen-etal-01}. 
We note here that, in contrast to the situation for $\Pb$, the total 
number of neutrons produced per kTon in the case of $\Fe$ is dominated 
by those produced through the NC interactions. Specifically, for 
$\Fe$, the NC 
induced neutrons constitute $\sim$ 62\% (71\%) of the total 
number of neutrons produced in the case of NO (IO) mass hierarchy. 
\subsection*{Comparison of $\Pb$ and $\Fe$ as detector materials}
Although in the above discussions we have not at all considered the 
issues of specific detector configurations and detection efficiencies of 
specific kinds of detectors, it is already clear from the results 
presented above that, per kTon of material, $\Fe$ gives the number of 
emitted neutrons lower by more than an order of magnitude compared to 
$\Pb$ both for CC and NC interactions. This is due to two reasons. 
Firstly, the total Gamow-Teller strength for $\nu_e$ CC cross section  
roughly scales as $N-Z$, $N$ and $Z$ being the number of neutrons and 
protons, respectively, in the nucleus.  
This neutron excess is 44 for $^{208}{\rm Pb}$ and only 4 for $^{56}{\rm 
Fe}$. Also, the neutron emission thresholds for both $^{208}{\rm Pb}$ 
and $^{208}{\rm Bi}$ are a few MeV lower compared to those of $^{56}{\rm 
Fe}$ and $^{56}{\rm Mn}$. As a result the contribution of most of the 
allowed strengths for both CC and NC cannot make any contribution for
$\Fe$ but they do for $\Pb$. Thus, as a detector material $\Pb$ is more 
efficient than $\Fe$ for observing supernova neutrinos. 

However, as clear 
from the discussions above, a $\Pb$ detector would be primarily 
sensitive to $\nue$'s through CC interactions, with neutrons 
produced through NC interactions of $\nue$, $\anue$, $\nux$ and 
$\anux$ comprising in total $\sim$ 20\% or less of all the produced 
neutrons. In contrast, in a $\Fe$ detector, most ($\gsim$ 60\%) of all 
the neutrons produced should be of NC origin. 
Thus, without the ability to 
separately identify the CC induced events (through identification of the 
accompanying $e^-$) in either detector, simultaneous detection of SN 
neutrinos in a $\Pb$ and a sufficiently large $\Fe$ detector, together 
with the knowledge of the expected fractions of total numbers of CC and 
NC events in each detector as discussed above, may, in principle, allow 
a determination of the 
fractions of $\mu$ and $\tau$ flavored neutrinos in the total SN 
neutrino flux. 

At the present time, however, neutrino cross sections on 
both iron and lead at the relevant neutrino energies are still rather 
uncertain, with no precise measurements of these cross sections being 
currently available, and so a proper assessment of the feasibility of 
the above approach will have to await the availability of more precisely 
determined neutrino cross sections on lead as well as iron. Also, 
realistic estimates of the minimum sizes of the lead and iron detectors 
that would allow extraction of meaningful information 
on the  $\numu$ and $\nutau$ components of the SN neutrino flux will 
require a more detailed analysis, than has been attempted here, 
involving the relevant statistical as 
well as systematic uncertainties --- the latter including those due 
to uncertainties in the relevant cross sections. 

\section{Summary and Conclusions}
\label{sec:summary}
We have presented the results of our study of the possibility of 
detecting SN neutrinos with $\Fe$ and $\Pb$ detectors through detection 
of neutrons emitted by the excited nuclei resulting from the interaction 
of SN neutrinos with the nuclei of these detector materials. In doing 
this, we have used the results of the most 
recent state-of-the-art Basel/Darmstadt simulations~\cite{Basel-Darmstadt-10}
for the supernova neutrino fluxes and average energies, which typically
yield closer fluxes among different neutrino flavors and lower average
energies compared to those given by earlier 
simulations~\cite{Totani-etal-98,Gava-etal-09}. Specifically, we have 
used the simulation results of the Basel/Darmstadt simulations for a 
$18\msun$ progenitor SN at a distance of 10 kpc. Our results for the 
numbers of neutron events per kTon of detector material are   
found to be significantly lower than those estimated in previous studies 
which were based on earlier simulations of SN neutrino emission. 
It will be of interest to study the implications 
of this result for the possibility of 
effectively distinguishing between the neutrino mass 
hierarchies using SN neutrino detection (see, e.g., 
Refs.~\cite{Vale-14,Vale-etal-16}). 

We also make the observation that, 
while $\Pb$ would be a better detector material 
than $\Fe$ in terms of the total number of neutrons produced per kTon of 
detector mass, $\sim$ 80\% or more of the produced 
neutrons in the case of 
$\Pb$ arise from CC interactions of $\nue$, whereas neutrons produced in 
$\Fe$ are dominated by those produced through NC interactions of all the 
six $\nu$ plus $\anu$ species. Thus, a sufficiently large $\Fe$ detector
--- large enough to compensate for the overall smaller $\nu\,$-$\Fe$ 
cross sections compared to $\nu\,$-$\Pb$ cross sections ---
can be a good NC detector for SN neutrinos. For example, the proposed 50
kTon iron calorimeter (ICAL)~\cite{ICAL-15} detector, though primarily
designed for studying neutrino properties using the relatively higher
energy (multi-GeV) atmospheric neutrinos, can also be a good NC
detector of SN neutrinos if it can be appropriately instrumented
with suitable neutron detectors. Thus, simultaneous 
detection of SN neutrinos in a $\Pb$ and a $\Fe$ detector can, in 
principle, provide an estimate of 
the relative fractions of the $\nue$ and the other five neutrino species 
in the SN neutrino flux, which would be a good probe of the 
SN neutrino production as well as flavor oscillation scenarios. 
It will be interesting to carry out more detailed analysis involving 
relevant statistical and systematic uncertainties --- the latter 
including those due to (currently somewhat large) uncertainties in 
neutrino cross sections on lead and iron --- to derive 
estimates of the minimum lead and iron detector sizes that would allow 
extraction of statistically significant information on the $\numu$ and 
$\nutau$ components of the SN neutrino flux. 

\noindent{\bf Acknowledgment:} One of us 
(SC) thanks Tobias Fischer for providing the numerical data for the 
temporal profiles of the luminosities and average energies of  
neutrinos of different flavors for the Basel/Darmstadt SN simulations 
used in this paper. SC also acknowledges partial support by the 
Deutsche Forschungsgemeinschaft through Grant No.~EXC 153 (``Excellence 
Cluster Universe") and by the European Union through 
the ``Initial Training Network Invisibles," Grant 
No.~PITN-GA-2011-289442.
%

\end{document}